\documentclass[letter]{aa} 
\usepackage{graphicx}
\usepackage{url}            
\usepackage{txfonts}
\usepackage{natbib}
\bibpunct{(}{)}{;}{a}{}{,}  
\usepackage{color}
\usepackage[percent]{overpic} 

\def\OII{[O\,\textsc{ii}]}

\begin{document}
   \title{
   The X-ray luminous galaxy cluster XMMU\,J1007.4+1237 at $z\!=\!1.56$\thanks{Based on observations under programme ID 081.A-0312 collected at the European Organisation for Astronomical Research in the Southern Hemisphere, Chile, and observations collected at the Centro Astron\'omico Hispano Alem\'an (CAHA) at Calar Alto, operated jointly by the Max-Planck Institut f\"ur Astronomie and the Instituto de Astrof\'isica de Andaluc\'ia (CSIC). }}

\subtitle{The dawn of  starburst activity in cluster cores}

   \author{
    R. Fassbender
          \inst{\ref{MPE}}
          \and
             A. Nastasi
          \inst{\ref{MPE}}
          \and
          H. B\"ohringer
          \inst{\ref{MPE}}
          \and
          R. \v{S}uhada
          \inst{\ref{MPE}}
       	  \and
          J.S. Santos
          \inst{\ref{Trieste}}
          \and
          P. Rosati
          \inst{\ref{ESO}}
          \and
          D. Pierini
          \inst{\ref{MPE}}
          \and
          M. M\"uhlegger
          \inst{\ref{MPE}}
           \and
          H. Quintana
          \inst{\ref{PUC}}
          \and
          A.D. Schwope
          \inst{\ref{AIP}}
          \and
          G. Lamer
          \inst{\ref{AIP}}
          \and
          A. de Hoon
          \inst{\ref{AIP}}
           \and
          J. Kohnert
          \inst{\ref{AIP}}
          \and
          G.W. Pratt
          \inst{\ref{Saclay}}
           \and
          J.J. Mohr
          \inst{\ref{MPE},\ref{USM},\ref{EC}} 
          }

  \institute{Max-Planck-Institut f\"ur extraterrestrische Physik (MPE),
              Giessenbachstrasse~1, 85748 Garching, Germany \\
              \email{rfassben@mpe.mpg.de} \label{MPE}
            \and
         INAF-Osservatorio Astronomico di Trieste, Via Tiepolo 11, 34131 Trieste, Italy \label{Trieste}
          \and
         European Southern Observatory (ESO), Karl-Scharzschild-Str.~2, 85748 Garching, Germany \label{ESO}
          \and
          Departamento de Astronom\'ia y Astrof\'isica, Pontificia Universidad
          Cat\'olica de Chile, Casilla 306, Santiago 22, Chile \label{PUC}
             \and
            Astrophysikalisches Institut Potsdam (AIP),
            An der Sternwarte~16, 14482 Potsdam, Germany \label{AIP}
           \and
            University Observatory Munich, Ludwigs-Maximillians University Munich,
            Scheinerstr. 1, 81679  Munich, Germany \label{USM}
         \and
        CEA \/ Saclay, Service d'Astrophysique, L'Orme des Merisiers, B\^at. 709, 91191 Gif-sur-Yvette Cedex, France \label{Saclay}
        \and
        Excellence Cluster Universe, Boltzmannstr.~2, 85748 Garching, Germany  \label{EC}
%
             }

   \date{Received November 18, 2010; accepted January 4, 2011}

 
  \abstract
   {Observational galaxy cluster studies at z$>$1.5 probe the formation of the first massive $M$$>$$10^{14}$\,M$_{\sun}$ dark matter halos, the early thermal history of the hot 
   ICM, and the emergence of the red-sequence population of quenched early-type galaxies.}
   {We present first results for the newly discovered X-ray luminous galaxy cluster XMMU\,J1007.4+1237 at $z\!=\!1.555$, detected and confirmed by the  XMM-{\it Newton}  Distant Cluster Project (XDCP) survey.}
   {We selected  the system as a serendipitous weak extended X-ray source in XMM-{\it Newton} archival data and followed it up with two-band near-infrared imaging and deep optical spectroscopy.}
   {We can establish XMMU\,J1007.4+1237 as a spectroscopically confirmed,  massive, {
   bona fide} 
   galaxy cluster with a bolometric X-ray luminosity of $L^{\mathrm{bol}}_{X,500}\simeq(2.1 \pm 0.4)\times 10^{44}$\,erg/s, 
   a red galaxy population centered on the X-ray emission, and a central radio-loud brightest cluster galaxy.  However, we see evidence for the first time that the massive end of the galaxy population and the cluster red-sequence are not yet fully in place. In particular, we find ongoing starburst activity for the third ranked galaxy close to the center and another slightly fainter object.   
   }
   {At a lookback time of 9.4\,Gyr, the cluster galaxy population appears to be caught in an important evolutionary phase, prior to full star-formation quenching and mass assembly in the core region. X-ray selection techniques are  an efficient means of identifying and probing the most distant clusters without any prior assumptions about their galaxy content.}



   \maketitle
%




\begin{figure*}[t]
    \centering
    \includegraphics[width=8.95cm, clip]{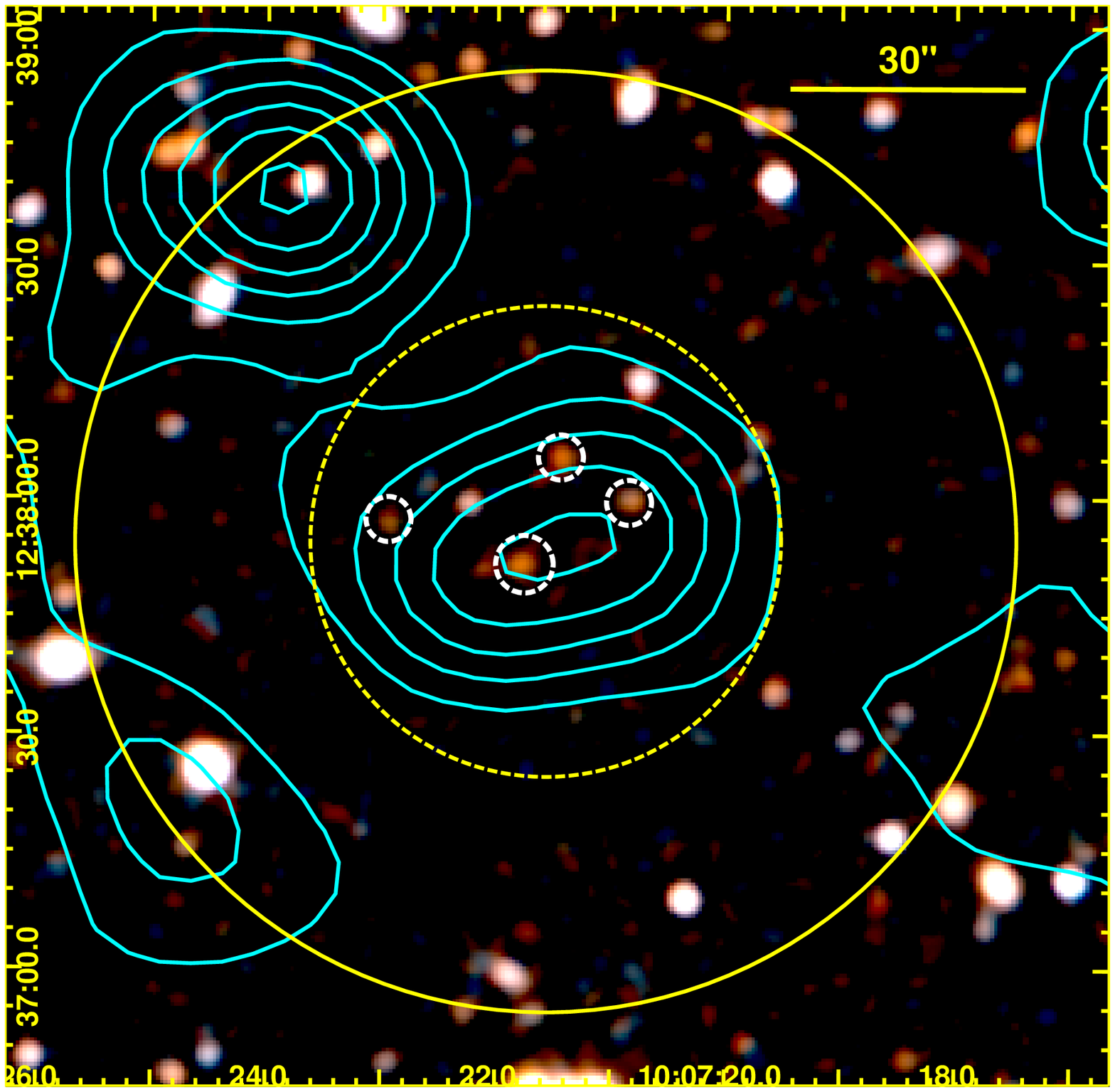}
   \hspace{0.1cm}
    \includegraphics[width=8.95cm, clip=true]{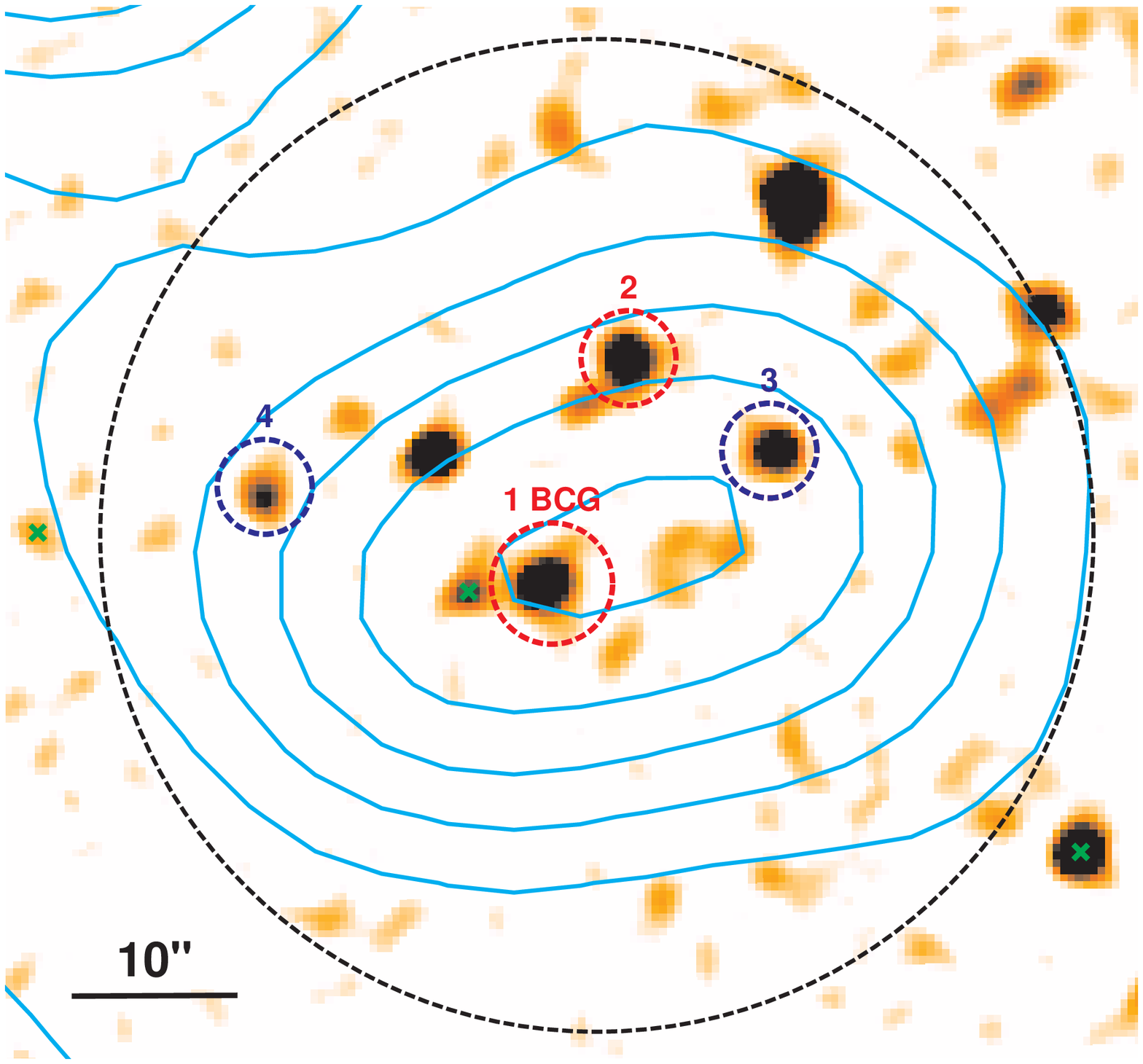}
      \caption{Optical/NIR appearance of the galaxy cluster XMMU\,J1007.4+1237 at z=1.555 with XMM-{\it Newton} X-ray contours overlaid in cyan  in the logarithmically spaced range [0.014, 0.074] counts arcsec$^{-2}$ (0.35-2.4\,keV).  {\em Left:} z+H-band color composite (2.3\arcmin$\times$2.3\arcmin) centered on the cluster. Spectroscopic cluster members are marked by 
      white circles,  30\arcsec \ (dashed) and 60\arcsec  \ (solid) radii around the X-ray center are indicated in yellow.
 {\em Right:} H-band zoom on the cluster core region. Red (IDs 1\,\&\,2) and blue starburst  (IDs 3\,\&\,4) cluster members are marked by small circles, spectroscopically confirmed foreground galaxies are indicated by green crosses, and the outer dashed circle represents the 30\arcsec \ radius.}
         \label{fig_OpticalAppearance}
\end{figure*}


\section{Introduction}

The `redshift desert' at $z\!>\!1.5$ has until recently also been known as a  `galaxy cluster desert', owing to the difficulty in detecting these systems at these redshifts and, in particular, the challenging spectroscopic cluster confirmation once the 4\,000\,\AA-break is redshifted to beyond 10\,000\,\AA.
\citet{Kurk2009a} reported on a proto-cluster-like structure at  $z\!=\!1.6$  in the deep GMASS spectroscopic survey without detectable extended X-ray emission in the Mega-second exposure of the Chandra Deep Field South.  \citet{Papovich2010a} and
\citet{Tanaka2010a}  independently confirmed an infrared (IR) selected  system at  $z\!=\!1.62$   in a deep survey field, 
and a subsequent detection of its weak  X-ray emission consistent with a total mass estimate in the group regime ($M_{200}\!\simeq\!6\!\times\!10^{13}\,\mathrm{M_{\sun}}$). Similar approximate system masses 
were  reported for the first X-ray selected cluster/group 
at $z\!=\!1.75$ in the 
XMM-{\it Newton} Lockman Hole pencil-beam 
survey 
by \citet{Henry2010a} and  the IR-selected system CL\,J1449+0856 at $z\!=\!2.07$  in the `Daddi Field' by \citet{Gobat2010a}, 
which subsequently showed indications of very weak extended X-ray emission.

Observational studies of galaxy clusters at the highest accessible redshifts provide a direct view of the early assembly phase of the most massive dark matter (DM) halos, the X-ray emitting hot intracluster medium (ICM), and their galaxy populations. In contrast to the dominating `red and dead' galaxies in local clusters with a well-defined red-sequence of quenched, passively evolving early-type galaxies (ETGs), dramatic changes are expected as the main star-formation epoch of the bulk of the cluster galaxies is approached. At $z\!>\!1.5$, corresponding to lookback times of 
$\ga$9.3\,Gyr, we are close to the global cosmic star formation (SF) density peak  \citep{HopkinsAM2006a}. At this epoch, hydrodynamical simulations suggest that the SF 
quenching effects in dense environments should vanish as the specific star formation densities in clusters approach the peak values in the field \citep[e.g.][]{Romeo2005a}. Recent studies of XMMXCS\,J2215.9-1738 at $z\!=\!1.46$ indeed found 
strong ongoing SF activity all the way to the central cluster core region \citep{Hayashi2010a,Hilton2010a} and for CL\,J1449+0856 at $z\!=\!2.07$ 
a tight cluster red-sequence 
cannot be discerned \citep{Gobat2010a}. However, the almost completely quenched star formation at the core of the very massive cluster XMMU\,J2235.3-2557 at $z\!=\!1.39$  \citep{Strazzullo2010a,Rosati2009a} supports the idea that there was accelerated galaxy evolution and early SF suppression in the more massive systems. 


Here we report on the discovery of a new {\it bona fide} galaxy cluster in the `redshift desert'. The system XMMU\,J1007.4+1237 at  $z\!=\!1.555$ was blindly (i.e. without optical/IR data) detected as a serendipitous extended X-ray source within the XMM-{\it Newton} Distant Cluster Project \citep[XDCP, ][]{HxB2005a,RF2007Phd}, a systematic 80\,deg$^2$ archival search for X-ray luminous systems at $z\!>\!0.8$. Section\,\ref{c2_Observations} of this Letter presents the X-ray data, near-infrared (NIR) imaging, and spectroscopic observations, followed by a discussion in
Sect.\,\ref{c3_Discussion} and conclusions in  Sect.\,\ref{c4_Conclusions}. 
We assume a $\Lambda$CDM concordance cosmology with ($H_0$, $\Omega_{\mathrm{m}}$, $\Omega_{\mathrm{DE}}$, w)=(70\,km\,s$^{-1}$Mpc$^{-1}$, 0.3, 0.7, -1).  
A redshift of $z\!=\!1.555$ corresponds to a lookback time of
9.40\,Gyr and 
 a projected physical scale of 8.47\,kpc/\arcsec. 


\section{Observations, data analysis, and results}
\label{c2_Observations}





\subsection{XMM-{\it Newton} X-ray selection  and {\it Chandra} cross-check}
\label{s2_XMM_Xray}


The original XDCP source detection run was applied to  a total of  470 XMM-{\it Newton} archival fields and performed with   \texttt{SAS} v6.5 using the tasks \texttt{eboxdetect} and \texttt{emldetect} for a sliding box detection with a subsequent maximum likelihood analysis.
The X-ray source associated with XMMU\,J1007.4+1237 was detected  at an off-axis angle of 10.7\arcmin \ in a medium deep observation of 22.2\,ksec (OBSID: 0140550601) targeting the AGN  PG\,1004+130. After applying a strict two-level cleaning process to remove periods  of enhanced solar flare activity, 18.0\,ksec (21.5\,ksec)  of data remained for the PN  (MOS) cameras. On the basis of  these data, the cluster was characterized as a compact extended source ($r_c\!\simeq\!7\arcsec$) with a 4\,$\sigma$ extent significance\footnote{ Corresponding to an extent parameter {\tt EXT\,ML=8.2} of the  maximum likelihood fitting task \texttt{emldetect}, which uses a single $\beta$-model with a fixed $\beta$=0.667 as a model profile for the source extent determination.} based on about 130 source counts (0.3-4.5\,keV).

To achieve a more accurate determination of source parameters,  we re-reduced the XMM-{\it Newton} data with  \texttt{SAS} v10.0.0 and applied 
the growth curve analysis (GCA) method of \citet{HxB2000a} in the soft 0.5-2\,keV band. We measure an unabsorbed  0.5-2\,keV band flux in the 
estimated $R_{500}$ aperture ($\simeq\!42$\arcsec) of  $f_{X,500}\simeq(5.6 \pm 1.1)\times 10^{-15}$\,erg\,s$^{-1}$\,cm$^{-2}$. Using the redshift information of Sect.\,\ref{s2_spectroscopy}, we find a luminosity of $L_{X,500}\simeq(8.6 \pm 1.7)\times 10^{43}$\,erg/s in the soft 0.5-2\,keV band, corresponding to a bolometric ICM energy output of  $L^{\mathrm{bol}}_{X,500}\simeq(2.1 \pm 0.4)\times 10^{44}$\,erg/s for the estimated $T_{X}$ quoted below.

Figure\,\ref{fig_OpticalAppearance}  shows the logarithmically spaced X-ray contours based on the adaptively smoothed X-ray surface brightness (SB) distribution of the cluster with significance levels spanning 2-12\,$\sigma$ above the mean background,  corresponding to 0.014-0.074 net source counts arcsec$^{-2}$ in the 0.35-2.4\,keV band.
Owing to the intrinsically compact nature and the faintness of the source, we cannot rule out residual  point-source contributions to the measured luminosity at this point. Under the assumption that the thermal ICM emission dominates, we estimate an X-ray luminosity-based total cluster mass  of   
$M^{\mathrm{est}}_{200}\!\simeq\!2\!\times\!10^{14}$\,M$_{\sun}$ following the approach of \citet{Fassbender2010a} and a corresponding ICM temperature of $T^{\mathrm{est}}_{X}\!\simeq\!4.3$\,keV using the M-T relation of \citet{Arnaud2005a}.


The 
target  PG\,1004+130 was also observed with {\it Chandra} for 41.6\,ksec, 
placing XMMU\,J1007.4+1237 at an off-axis angle of 11\arcmin. 
After a standard reduction, we could find the cluster source with  a total of about 80 counts 
within $R_{500}$ with both {\it celldetect} (2.8\,$\sigma$) and {\it wavdetect} (5.6\,$\sigma$). However, at this off-axis angle the {\it Chandra} PSF ($\simeq$6\arcsec) is only marginally better than the XMM-{\it Newton} resolution, thus allowing  only a consistency check at this point. Figure\,\ref{fig_HbandDensities} shows the log-spaced Chandra SB contours of the adaptively smoothed 0.7-2\,keV band image  in blue 
 covering a dynamic range of  0.014-0.082 source counts arcsec$^{-2}$.   
We can confirm the principal cluster extent direction along the east-west axis and find a more pronounced 
main flux peak in close proximity to the brightest cluster galaxy (BCG).

\subsection{Near-infrared follow-up imaging}
\label{s2_imaging}

For the optical counterpart  identification 
we obtained medium deep H and z-band  imaging data with the prime-focus wide-field (15.4\arcmin$\times$15.4\arcmin) near-infrared  (NIR)  camera OMEGA2000 
at the Calar Alto 3.5m telescope. First observations  were performed on 4 January 2006 in the H-band (50\,min) in moderate to poor conditions with 1.1-1.6\arcsec \ seeing, complemented by additional data in photometric conditions on 7  January 2007 in H  (30\,min) and z-band (50\,min) with 0.9-1.3\arcsec \ seeing. The data were reduced with a designated OMEGA2000 NIR pipeline  \citep{RF2007Phd}  and co-added to deep image stacks after inspection of all individual frames. 
The final stacks comprise 63\,min of clean data in H (1.39\arcsec \ FWHM) and  28\,min in z (1.17\arcsec \ FWHM), after rejection of frames with low atmospheric transparency in H (12\%) and corrupted frames in z (44\%). This results in 50\%-completeness limits (Vega) of $H_\mathrm{lim}\!\sim\!20.7$ and $z_\mathrm{lim}\!\sim\!22.5$. 
Dual-image band photometry with the H stack as a detection frame was performed with \texttt{Sextractor} \citep{Bertin1996a}. The photometric calibration was tied to 2MASS point sources \citep{Cutri2003a} in H and designated 
standard star observations in z  \citep{Smith2002a}, which was cross-checked with SDSS photometry in the science field. 

\begin{figure}[t]
   \centering
    \includegraphics[width=9cm, clip=true]{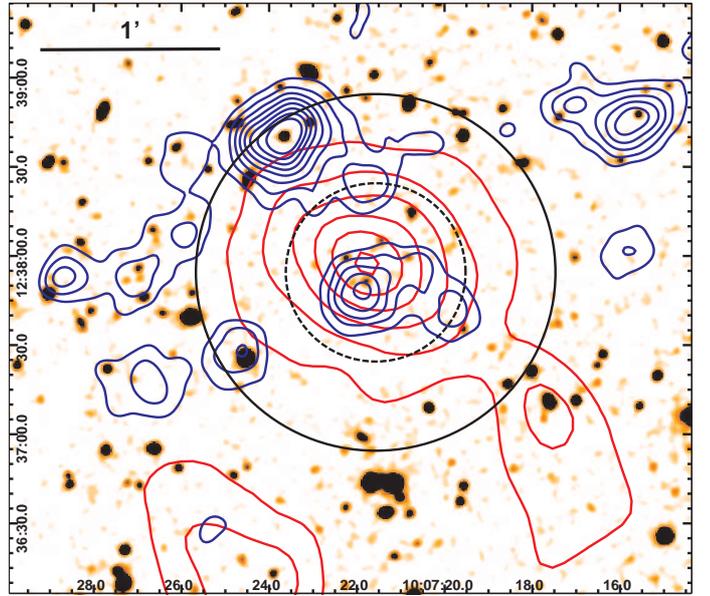}
      \caption{H-band view of the 3.8\arcmin$\times$3.3\arcmin \ cluster environment. 
    Red contours indicate the log-spaced projected densities of very red galaxies with colors 3$\le$z$-$H$\le$4 (see Fig.\,\ref{fig_CMD})  spanning the levels 2-16 arcmin$^{-2}$, Chandra X-ray surface brightness contours (0.7-2\,keV, log-spaced) are shown in blue  in the range 0.014-0.082 counts arcsec$^{-2}$.  30\arcsec \ (dashed) and 60\arcsec  \ (solid) radii around the X-ray center are marked by  circles. 
   }
         \label{fig_HbandDensities}
\end{figure}

\begin{figure}[t]
   \centering
     \includegraphics[width=9cm, clip=true]{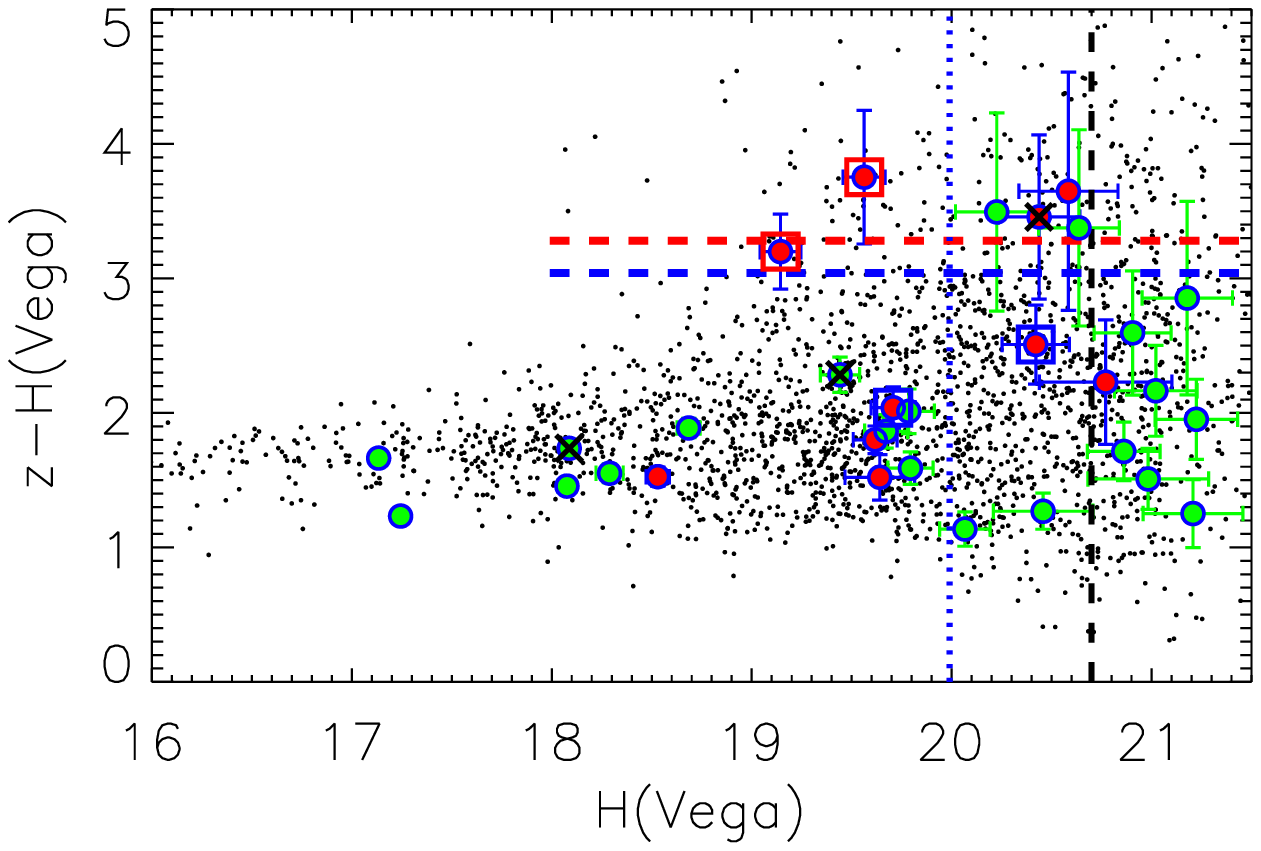}
     \includegraphics[width=9cm, clip=true]{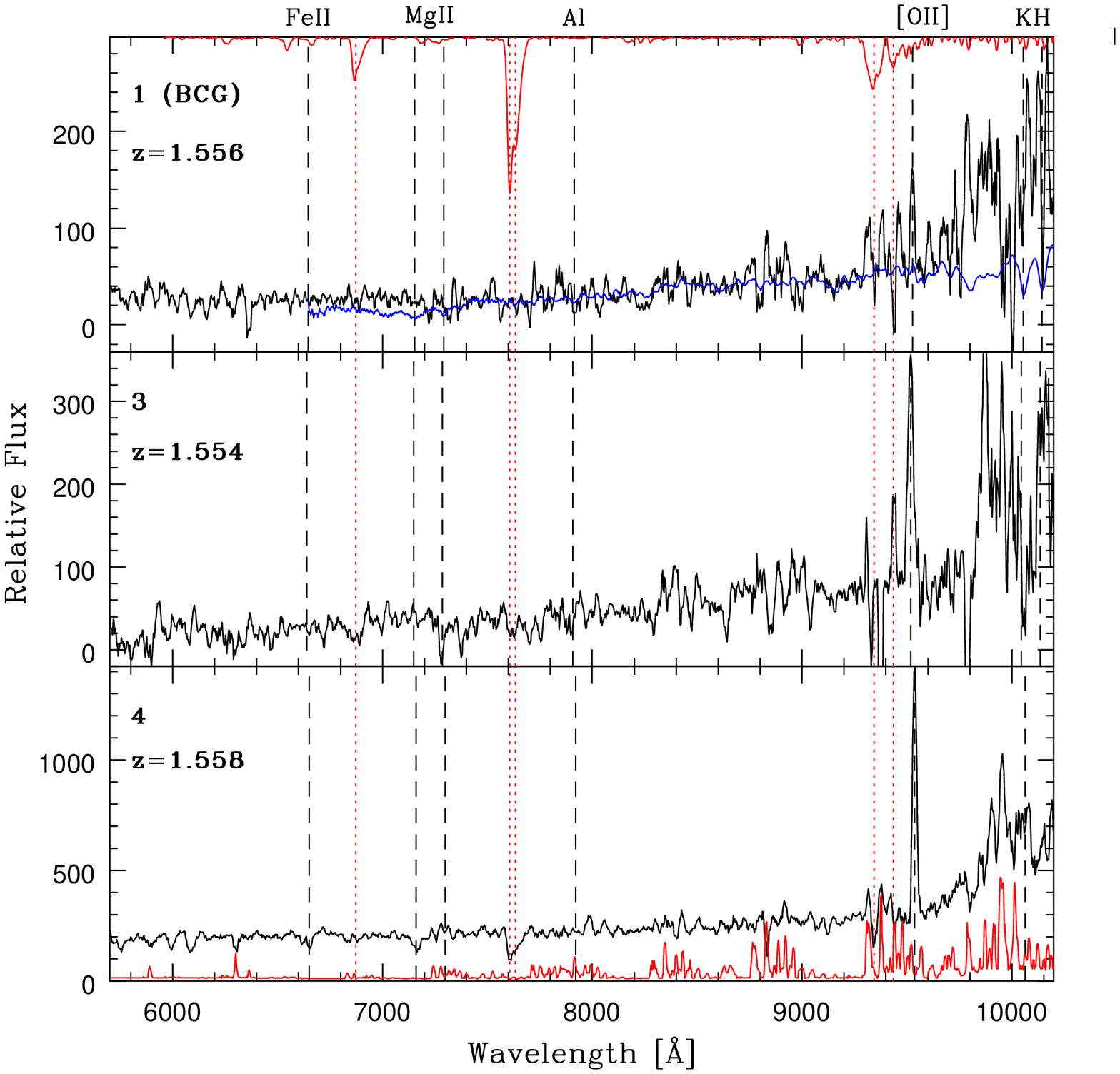}
      \caption{ {\em Top:}
      z$-$H vs. H CMD 
      of the cluster field. Galaxies with projected cluster-centric distances of $<$30\arcsec \ (30-60\arcsec) are displayed in red (green). Secure spectroscopic cluster members are marked by open squares, foreground objects are crossed-out. The vertical 
      blue line indicates the apparent H magnitude of an L* galaxy at $z\!=\!1.56$, the black 
      line shows the H-band magnitude limit.
 Two solar metallicity SSP models for 
  formation redshifts of $z_f\!=\!5$ (red dashed line) and $z_f\!=\!3$ (blue dashed) are overplotted. 
       {\em Bottom:}  VLT/FORS\,2 spectra of three secure cluster members smoothed with a 7 pixel boxcar filter, associated 
   IDs (1,3,4) refer to Table\,\ref{tab_specmembers}. Redshifted spectral features are indicated together with an overlaid LRG template spectrum in blue (ID 1), sky emission lines (bottom)  and telluric absorption (top) are shown in red.  
       } \label{fig_CMD}
\end{figure}

Figure\,\ref{fig_OpticalAppearance}  shows the z+H color composite image\footnote{ Using z for the blue channel,  z+H for green, and H for red.} of the cluster (left) and a H-band zoom on the core region  (right)  with the  XMM-{\it Newton}  X-ray contours overlaid (Sect.\,\ref{s2_XMM_Xray}). The larger-scale cluster environment with surface density contours of very red galaxies   in the range 2-16\,arcmin$^{-2}$  is displayed in Fig.\,\ref{fig_HbandDensities} (red). The z$-$H versus H color magnitude diagram (CMD) of the field is presented in the top panel of Fig.\,\ref{fig_CMD}. 
Although the 
current photometric depth is limited, reaching m*+0.7 in the detection band,  a fair initial redshift estimate for the cluster of $z\!\simeq\!1.7\!\pm\!0.2$  \citep{RF2007Phd} could still be obtained prior to spectroscopy based on the comparison of the observed color of the reddest 
central objects 
with simple stellar population (SSP) 
models (Fig.\,\ref{fig_CMD}, horizontal lines).

\subsection{Spectroscopic confirmation and radio  properties}
\label{s2_spectroscopy}

For the final spectroscopic confirmation of the cluster, we obtained deep observations  with VLT/FORS\,2 (Program ID: 081.A-0312) using a single MXU-mode (Mask eXchange Unit) slit-mask (field-of-view 6.8\arcmin$\times$6.8\arcmin). A total of five 1\,h observing blocks were executed targeting XMMU\,J1007.4+1237 in good seeing conditions of 0.8\arcsec \ on 25 April 2008 and 5 May 2008. The chosen instrument setup with the 300\,I grism without blocking filter and a slit width of 1\arcsec\, provides a wavelength coverage of 5\,800--10\,500\,\AA \ with a resolution of $R\!=\!660$. The resulting ten individual frames with 21.8\,min net exposure times each (3.6\,h in total) were reduced with a new FORS\,2 adaptation of the 
{\it VIMOS Interactive Pipeline and Graphical Interface} \citep[VIPGI, ][]{Scodeggio2005a}, which includes bias subtraction, flat field corrections, image stacking, extraction of background-subtracted 1D spectra, and wavelength calibration by means of a helium-argon lamp reference line spectrum (Nastasi et al., in prep.). The final stacked spectra  
were corrected for the sensitivity function of the  FORS\,2  instrument and then cross-correlated with a  galaxy template library for a semi-automated redshift determination using the {\it IRAF} package  {\it RVSAO}  \citep{Kurtz1998a} and  {\it EZ}  \citep{Garilli2010a}.

Owing to the faintness of the sources and the limited photometric information, the primary slit targets were red galaxies (z$-$H$>$2) within the spectroscopic magnitude limit and close to 
the X-ray center.  
Indeed, the four brightest targeted central galaxies at d$_{\mathrm{cen}}$$<$25\arcsec, marked in Fig.\,\ref{fig_OpticalAppearance}, were found at redshifts 
1.53$\le$z$\le$1.56 (Fig.\,\ref{fig_CMD}  and Table\,\ref{tab_specmembers}). The existence of extremely strong \OII \ emission lines in  the galaxy spectra with IDs 3\,\&\,4 allowed a straightforward determination of concordant redshifts at 1.554 and 1.558, respectively. 
The redshift determination of passive galaxies 
at $z\!>\!1.5$ is close to the feasibility limit of  
FORS\,2 with reasonable exposure times. However,  for the BCG (ID 1) we were able to identify a  weak 
  \OII \ line and the 4\,000\,\AA-break at the limit of the wavelength coverage, to obtain a redshift of 1.556. The assigned redshift of 1.530 
for the reddest galaxy  (ID 2) is mainly based on a  weak 
 \OII \ line, and it therefore has a lower redshift confidence level.
%
Based on the available spectroscopic information for XMMU\,J1007.4+1237, 
we assign a median system  redshift of $z\!=\!1.555\!\pm\!0.003$. 
In the absence of a detailed velocity structure measurement, 
we treat the very red galaxy with ID 2 as a tentative member with a preliminary rest-frame velocity offset of about  $\Delta v\!=\!-2900$\,km\,s$^{-1}$.

To obtain a first assessment of the approximate 
ongoing star-formation activity in the member galaxies of XMMU\,J1007.4+1237, we measured the  \OII \ equivalent widths and applied the  \OII-SF relation 
of \citet{Kennicutt1992a}, which provides first-order SF rates
corrected for an average dust extinction 
in (local) star-forming systems 
(see Table\,\ref{tab_specmembers}). 
This first estimate yields SF rates for the blue galaxies with IDs 3\,\&\,4 of about 75 and 31\,M$_{\sun}$\,yr$^{-1}$, and roughly 11 and 5\,M$_{\sun}$\,yr$^{-1}$ for the red members with IDs 1\,\&\,2, respectively. 
However, these crude values could be significantly underestimated 
given the large uncertainties related to the dust extinction properties of high-z galaxies  \citep[e.g.][]{Calzetti2000a,Kewley2004a,Pierini2005a} and our limited photometric coverage.

\label{s2_radio}

The cluster center of XMMU\,J1007.4+1237 contains a 1.4\,GHz radio source with a flux of 2.1$\pm$0.4\,mJy identified by the  NVSS survey \citep{Condon1998a}.  This radio source, NVSS\,J100721+123749, is unresolved at the resolution level  of the survey (45\arcsec). The brightest object within the positional radio source uncertainty of  8\arcsec \ is the BCG at a distance of  4\arcsec, which is hence the most probable optical counterpart.  In this case, the measured flux translates into a total radio power of $P_{1440\,\mathrm{MHz}}\!\simeq\!3\!\times\!10^{25}$ W\,Hz$^{-1}$ at the cluster redshift  under the assumption of a typical spectral index in the common range [-1, -0.7] \citep[e.g.][]{Miley2008a}.

\section{Discussion }
\label{c3_Discussion}

In the local Universe, the likely association of the BCG with a powerful radio source would classify 
XMMU\,J1007.4+1237 
as a  probable cool core cluster (CCC) since the radio power scales inversely with the central cooling time \citep{Mittal2009a}. Additional support for a possible CCC scenario arises from the BCG location in close proximity to the center of the galaxy distribution and the extended X-ray emission, and in particular the very peaked X-ray surface brightness 
\citep{Sanderson2009a}. However, at a lookback time of 9.4\,Gyr  these CCC correlations are not yet established and the central radio source might still be associated with an X-ray AGN \citep{Hickox2009a}, hence point-like central X-ray emission.

Extended, non-thermal X-ray lobes  associated with  high-z radio galaxies have been reported for a number of $z\!>\!1$ systems
\citep[e.g.][]{Fabian2003b,Erlund2008a,Fabian2009a,Alexis2010a}. These sources with an inverse Compton (IC) origin of their diffuse emission can in principle contribute to the measured flux or even mimic the extended thermal ICM emission characteristic for well developed clusters. 
A contribution of  IC X-ray emission to that of XMMU\,J1007.4+1237 cannot presently be ruled out or constrained as this would require  currently unavailable high-resolution radio observations plus very deep {\it Chandra} and XMM-{\it Newton} data. 
However, the available limited X-ray spectral information on XMMU\,J1007.4+1237 is consistent with a soft thermal spectrum and the observed overall morphological elongation is expected at this redshift by structure formation simulations  \citep[e.g.][]{Springel2001a,Boylan2009a}. Moreover, a very compact central core of the surface brightness distribution eases the X-ray selection and may be the signature of an evolved high-z system \citep{Santos2008a,Rosati2009a}.


With  the thermal ICM assumption and the derived  
X-ray-luminosity-based total mass estimate of M$^{\mathrm{est}}_{200}\!\simeq\!2\!\times\!10^{14}$\,M$_{\sun}$,   XMMU\,J1007.4+1237 is currently the most massive known {\it bona fide} galaxy cluster in the `redshift desert'. We note that 
the goodness of the mass estimate at this distance depends more critically on the exact redshift evolution of the L-M scaling relation than the accurate X-ray luminosity, as long as the potential  non-thermal  
contribution is sub-dominant. The applied L-M relation of  \citet{Fassbender2010a} assumes a slower redshift evolution than self-similar model predictions (i.e. higher mass for a given $L_{\mathrm{X}}$),
fully consistent with the latest data compilation of Reichert et al. (in prep.) and recent simulations including preheating \citep{Stanek2010a}.


Even though the spatial density peak of red galaxies  (Fig.\,\ref{fig_HbandDensities}) coincides with the detected X-ray emission, the color-magnitude diagram (Fig.\,\ref{fig_CMD}, top) markedly differs from lower-z cluster galaxy populations. While the z$-$H colors of the BCG and the second ranked galaxy (red squares in Fig.\,\ref{fig_CMD}) are consistent with SSP model predictions for a high stellar formation redshift ($z_f\!\simeq\!5$), the total H-band magnitude of the BCG (H*-0.8) with respect to the expected characteristic luminosity 
H* at this redshift is more than one magnitude fainter than typical counterparts at z$<$1 \citep[e.g.][]{Smith2010a}, indicating that a significant fraction of the stellar BCG mass has not  yet been assembled.  

The third ranked, super-L* galaxy (H*-0.3) 
(left blue square and ID 3 in Fig.\,\ref{fig_CMD}), located at a projected cluster-centric distance of only 100\,kpc, is found to be a blue starburst galaxy. A second 
one  (right blue square, ID 4) was identified at a fainter magnitude (H*+0.4) at  r$_{\mathrm{proj}}\!\simeq\!170$\,kpc. At this intermediate magnitude range around H*$\simeq$20\,mag, the CMD is almost devoid of objects close to the expected color of red-sequence galaxies (red dashed line), in stark contrast to studies of massive early-type galaxies in clusters at z$<$1.4 \citep[e.g.][]{Lidman2008a,Mei2009a,Lerchster2010a}.
Hence, the cluster red-sequence in this $z\!=\!1.555$ system does not yet seem to be fully in place, as some of the brightest central cluster member galaxies are still caught at an epoch of strong starburst activity.

\section{Summary and conclusions}
\label{c4_Conclusions}

We summarize our findings as follows (see also Table\,\ref{table_results}):

   \begin{enumerate}
      \item We have presented first results from our study on the newly discovered {\it bona fide} galaxy cluster XMMU\,J1007.4+1237 at $z\!=\!1.555$, which was blindly selected as a serendipitous  extended X-ray source within the XMM-{\it Newton} Distant Cluster Project (XDCP) with a subsequent identification of a coincident red galaxy population and a central BCG. 
      \item On the basis of a measured bolometric X-ray luminosity of 
      $2.1\!\times\!10^{44}$\,erg\,s$^{-1}$\,($\pm$20\%), we estimated a total system mass of M$^{\mathrm{est}}_{200}\!\sim\!2\!\times\!10^{14}$\,M$_{\sun}$, making XMMU\,J1007.4+1237 the most massive currently known confirmed cluster in the `redshift desert' at z$>$1.5.  
      \item The cluster redshift could be established based on two central starburst galaxies and the BCG within $\Delta$z=0.004 using deep optical VLT/FORS\,2 spectroscopy. A fourth associated, very red galaxy was confirmed at a rest-frame velocity offset of about $-2900$\,km/sec. 
        \item  The central brightest cluster galaxy is located in close proximity to the X-ray peak and the maximum  density of very red galaxies, and is most likely the origin of the system's strong 1.4\,GHz radio emission with   P$_{\mathrm{1440\,MHz}}\!\simeq\!3\!\times\!10^{25}$\,W\,Hz$^{-1}$. While the BCG color is consistent with a large stellar formation redshift ($z_f\!\simeq\!5$),  the total H-band luminosity is found to be more than one magnitude fainter than typical lower-z systems.        
         \item 
      The cluster red-sequence of XMMU\,J1007.4+1237 does not appear to be not fully established yet. In particular, the luminosity range close to the characteristic magnitude H* seems to be deficient in red galaxies. Consistently, the third-ranked  member galaxy close to the cluster center was still caught at an epoch of strong starburst activity.  
      \end{enumerate}

\noindent
At a lookback time of 9.4\,Gyr, we seem to have reached the cosmic epoch where strong evolutionary effects  are still shaping the massive end of the cluster galaxy population. Upcoming deep multiwavelength observations of XMMU\,J1007.4+1237 promise to 
establish a more  precise picture of the cluster constituents at this important redshift.


\begin{table*}[t]    
\caption{Spectroscopic cluster members of XMMU\,J1007.4+1237. Listed are total Vega H-band magnitudes, z$-$H colors, projected cluster-centric distances  in arcseconds and kpc relative to the X-ray center (see Table\,\ref{table_results}), and 
first estimates of the ongoing SF activity following \citet{Kennicutt1992a}.
} \label{tab_specmembers}

\centering
\begin{tabular}{ c c c c c c c l c l}
\hline \hline

ID & RA     & DEC      & H                 & z$-$H        & d$_{\mathrm{cen}}$  & r$_{\mathrm{proj}}$  & SFR                     & $z_{\mathrm{spec}}$ & Comment \\

   &  J2000 &  J2000  &  mag  & mag  &  \arcsec      & kpc              & M$_{\sun}$\,yr$^{-1}$    &                   \\

\hline

1  & 10:07:21.8 & +12:37:51.5  &  19.2 &  3.2   &  4.3  & 37 & 11  & 1.556 & BCG  \\
2  & 10:07:21.4 & +12:38:05.1  & 19.6  & 3.8   &  10.9  & 92 & 5  & 1.530  & $\Delta v\!=\!-2900$\,km\,s$^{-1}$ \\
3  & 10:07:20.8 & +12:37:59.6  &  19.7 & 2.1   & 12.3  & 104 & 75 & 1.554 & strong \OII  \\
4  & 10:07:22.9 & +12:37:56.7  & 20.4  & 2.6   & 20.8  & 176 & 31  & 1.558 & strong \OII     \\

\hline
\end{tabular}
\end{table*}

\begin{table}[h]    
\caption{Properties of the galaxy cluster XMMU\,J1007.4+1237.} \label{table_results}

\centering
\begin{tabular}{ l l l}
\hline \hline

Property & Value & Unit   \\
\hline

RA      &  10:07:21.6       &    \\
DEC     & +12:37:54.3  &     \\
N$_{\mathrm{H}}$ & $ 3.59 \times 10^{20}$   &  cm$^{-2}$      \\
z       &  $1.555\pm 0.003$    &      \\

\hline

f$^{0.5-2\,\mathrm{keV}}_{\mathrm{X,500}}$ & $ (5.6 \pm 1.1)\times 10^{-15}$ & erg\,s$^{-1}$\,cm$^{-2}$   \\
L$^{0.5-2\,\mathrm{keV}}_{\mathrm{X,500}}$  & $(0.86 \pm 0.17) \times 10^{44}$ & erg\,s$^{-1}$   \\
L$^{\mathrm{bol}}_{\mathrm{X,500}}$  & $(2.1 \pm 0.4) \times 10^{44}$ & erg\,s$^{-1}$  \\
M$^{\mathrm{est}}_{200}$ &   $\sim$$2 \times 10^{14}$  & M$_{\sun}$       \\

\hline

P$_{\mathrm{1440\,MHz}}$  & $ 3 \times 10^{25}$ & W\,Hz$^{-1}$  \\

\hline
\end{tabular}
\end{table}

\begin{acknowledgements}
We acknowledge the excellent support provided by Calar Alto and VLT staff in carrying out the service observations. 
This research was supported by the DFG cluster of excellence ÒOrigin and Structure of the UniverseÓ (www.universe-cluster.de), by
the DFG under grants Schw536/24-1, Schw 536/24-2, BO 702/16-3, and the German DLR under grant 50 QR 0802. 
RF acknowledges the hospitality of the Department of Astronomy and Astrophysics at Pontificia Universidad  Cat\'olica de Chile. 
HQ thanks the FONDAP Centro de Astrofisica for partial support.
The XMM-Newton project is an ESA Science Mission with instruments and contributions directly funded by ESA Member
States and the USA (NASA). 
This research has made use of the NASA/IPAC Extragalactic Database (NED) which is operated by the Jet Propulsion Laboratory, California Institute of Technology, under contract with the National Aeronautics and Space Administration. 
\end{acknowledgements}


\bibliographystyle{aa} 
\bibliography{../../BIB/RF_BIB_10}


\end{document}